\documentclass[superscriptaddress,twocolumn,prl,showpacs,floatfix]{revtex4}
\usepackage[dvips]{graphicx}
\usepackage{amsmath,amsfonts,amssymb,bm,ulem}
\usepackage{graphicx}

\begin{document}

\title{Magnetoresistivity in a Tilted Magnetic Field in
p-Si/SiGe/Si Heterostructures with an Anisotropic g-Factor: Part II.}

\author{I.L.~Drichko}
\affiliation{A.F. Ioffe Physical-Technical Institute of Russian
Academy of Sciences, 194021 St. Petersburg, Russia}
\author{I.Yu.~Smirnov}
 \affiliation{A.F. Ioffe
Physical-Technical Institute of Russian Academy of Sciences, 194021
St. Petersburg, Russia}
\author{A.V.~Suslov}
\affiliation{National High Magnetic Field Laboratory, Tallahassee,
FL 32310, USA}
\author{O. A. Mironov}
\affiliation{Warwick SEMINANO R\&D Centre, University of Warwick
Science Park, Coventry CV4 7EZ, UK}
\author{D.R. Leadley}
\affiliation{Department of Physics, University of Warwick, Coventry
CV4 7AL, UK}

\date{\today}

\begin{abstract}
The magnetoresistance components  $\rho_{xx}$ and  $\rho_{xy}$ were
measured in two p-Si/SiGe/Si quantum wells that have an anisotropic
g-factor in a tilted magnetic field as a function of temperature,
field and tilt angle. Activation energy measurements demonstrate the
existence of a ferromagnetic-paramagnetic (F-P) transition for a
sample with a hole density of $p$=2$\times$10$^{11}$\,cm$^{-2}$.
This transition is due to crossing of the 0$\uparrow$ and
1$\downarrow$  Landau levels. However, in another sample, with
$p$=7.2$\times$10$^{10}$\,cm$^{-2}$, the 0$\uparrow$ and
1$\downarrow$ Landau levels coincide for angles
$\Theta$=0-70$^{\text{o}}$. Only for $\Theta
>$70$^{\text{o}}$ do the levels start to diverge which, in turn, results in the
energy gap opening.

\end{abstract}

\pacs{73.23.-b, 73.43.Qt}

\maketitle

\newpage

\section{Introduction} \label{intr}

Magnetotransport measurements on dilute p-Si/SiGe/Si structures,
with two-dimensional hole gas (2DHG) densities of about
10$^{11}$\,cm$^{-2}$, have revealed an unusual phenomenon at filling
factor $\nu$=3/2, the so-called "re-entrant" metal-insulator
transition.~\cite{1,2,3,4,5,6} This phenomenon manifests itself as
an additional peak of the magnetoresistance $\rho_{xx}(T, \Theta)$
at $\nu$=3/2. The peak demonstrates an insulator type behavior, i.e.
its magnitude increases with decreasing sample
temperature.~\cite{3,5}

The authors of Ref.~\onlinecite{2} explained this appearance by the
presence of smooth long-range potential fluctuations having a
magnitude comparable to the Fermi energy.  However, in
Refs.~\onlinecite{3,4,5} the magnetoresistance anomaly was
attributed to a crossing of Landau levels (LLs) with different spin
directions 0$\uparrow$ and 1$\downarrow$ as the magnetic field
increased.  It appears that some p-Si/SiGe/Si systems show a
magnetoresistance anomaly at $\nu=3/2$ that depends on the tilt
angle between the magnetic field and sample normal,~\cite{6} whereas
in other p-Si/SiGe/Si systems this anomaly is not manifested at
all.~\cite{4} A third set of p-Si/SiGe/Si systems have such anomaly
in $\rho_{xx}$ at $\nu=3/2$, but it does not depend on the tilt
angle.~\cite{3}

In our earlier article~\cite{7}, we analyzed \textit{the
conductivity} at $\nu$=2 in tilted magnetic fields in a sample with
$p$=2$\times$10$^{11}$\,cm$^{-2}$ and demonstrated the presence of a
ferromagnetic-paramagnetic (F-P) transition at a tilt angle of about
60$^{\text{o}}$.  It should be noted that at $\nu$=3/2 we did not
observe any significant variation of the conductivity, instead a
resistivity peak of the re-entrant-transition-type occurred in this
region of filling factor.  We therefore focused our research on the
$\nu$=2 region, i.e. in the vicinity of the
ferromagnetic-paramagnetic transition.  The magnetoresistance
components  $\rho_{xx}$ and  $\rho_{xy}$ for the p-Si/SiGe/Si
structure were measured in a tilted magnetic field, from which the
conductivity $\sigma_{xx}$ was calculated together with its
dependence on temperature $T$, magnetic field, and the tilt angle
$\Theta$. Such an approach allowed us to approximately calculate
values of the Landau level energies, rather than just providing a
qualitative description of the phenomenon, as was presented in
Refs.~\onlinecite{1,2,3,4,5,6}.  The F-P phase transition seen at
$\nu\cong$2, $T$=0.3 K, and  $\Theta \approx$60$^{\text{o}}$, is the
result of crossing of the 0$\uparrow$ and 1$\downarrow$ LLs. This
transition is characterized by a jump in the filling factor and by a
coexistence of both phases in the transition region. A F-P
transition has previously been reported in p-Si/SiGe/Si at $\nu$=4
and 6 in a tilted magnetic field by the authors of
Ref.~\onlinecite{8}.

The present paper is an continuation of our previous
article~[\onlinecite{7}] and has three aims: (i) to study the
dependence of the energy gap between LLs 0$\uparrow$ and
1$\downarrow$ on the magnetic field tilt angle $\Theta$ to provide
further confirmation of the crossing of these levels, in the
p-Si/SiGe/Si sample with $p$=2$\times$10$^{11}$\,cm$^{-2}$; (ii) to
investigate the conductivity anisotropy in this sample, by measuring
the conductivity at different orientations of the magnetic field
component in the sample plane with respect to the current:
$B_{\parallel}
\parallel I$ and $B_{\parallel}
\perp I$ , and comparing this with the theoretical model proposed in
~[\onlinecite{9}]; (iii) to measure the magnetoresistance in a
tilted magnetic field for another p-Si/SiGe/Si sample with a lower
density of $p$=7.2$\times$10$^{10}$\,cm$^{-2}$ and compare it with
the experimental data obtained by other groups on similar samples
~\cite{3,4,6}, with the hope of clearing up the inconsistency of the
previous results mentioned above.

\section{Experiment and Discussion} \label{exp}
In this research we studied two p-Si/SiGe/Si systems grown on a Si
(100) substrate that consisted of a 300 nm Si buffer layer followed
by a 30 nm Si$_{(1-x)}$Ge$_{x}$ layer, 20 nm undoped Si spacer, and
50 nm layer of B-doped Si with a doping concentration of
2.5$\times$10$^{18}$\,cm$^{-3}$. One sample had x=0.08, yielding
$p$=7.2$\times$10$^{10} $\,cm$^{-2}$, and the second had x=0.13,
with $p$=2$\times$10$^{11}$\,cm$^{-2}$. Both samples had a hole
mobility of about 1$\times$10$^4$\,cm$^2$/Vs at liquid-helium
temperatures.

\begin{figure}[h]
\centerline{
\includegraphics[width=8cm,clip=]{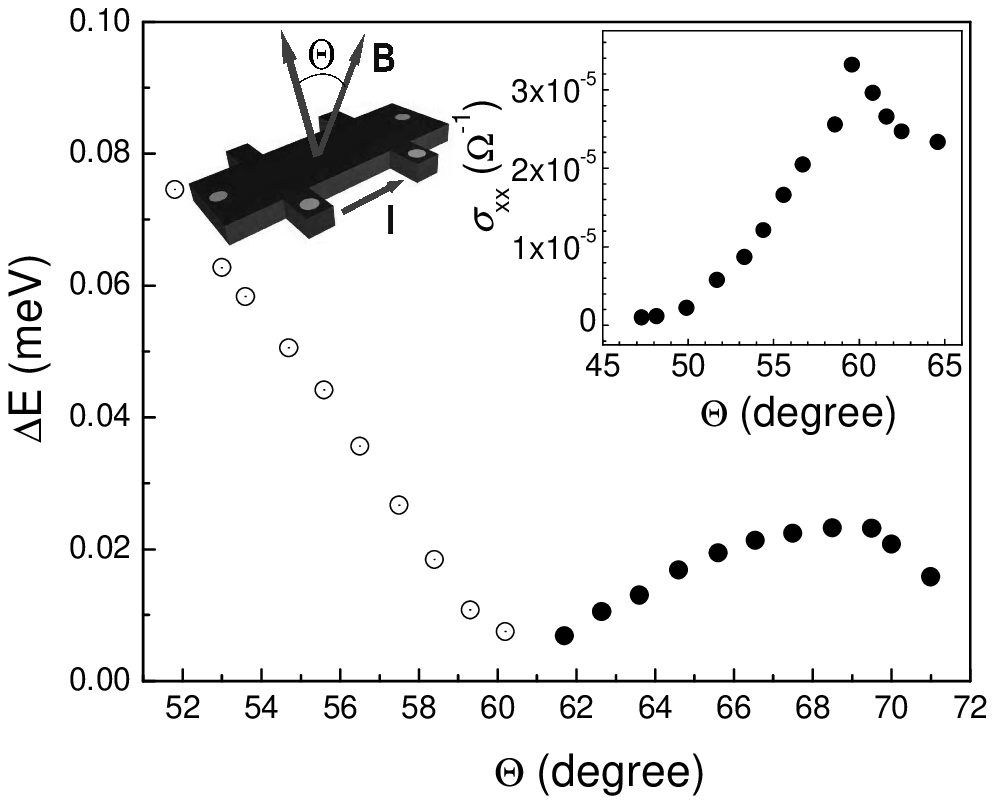}
} \caption{Dependence of the activation energy on tilt angle
$\Theta$. Inset: Dependence of the conductivity $\sigma_{xx}$ on
$\Theta$ at $\nu\approx$2; $T$=0.3 K. \label{Activa}}
\end{figure}

In the sample with $p$=2$\times$10$^{11}$\,cm$^{-2}$ we measured the
temperature dependence of the conductivity at different tilt angles
$\Theta$ over the temperature range 20 mK to 1 K, from which we were
able to determine the activation energy $\Delta E$ at various angles
via the slope of the Arrhenius curves: $\ln \sigma_{xx} \propto 1 /
T$.  The dependence of the activation energy on the tilt angle
$\Theta$ is shown in Figure~\ref{Activa}, where it can clearly be
seen that the activation energy achieves a minimum at $\Theta
\approx$60$^{\text{o}}$.  The conductivity $\sigma_{xx}$($\Theta$)
at the minima of oscillations at $\nu\cong$2, also shows a maximum
as a function of tilt angle at $\Theta \approx$60$^{\text{o}}$, as
shown in the inset to Figure~\ref{Activa}.

It is worth noting that when the measurements are performed with the
magnetic field normal to the sample plane the energy gap related to
$\nu \cong$ 2 is about 3.2 K (0.28 meV).  Thus, we are justified in
extracting the energy gap value from the temperature range of 200 mK
- 1 K. When the tilt angle approaches 60$^{\text{o}}$ the size of
the energy gap is very small, due to the LLs crossing. So, whilst
the actual gap value obtained here is subject to considerable
uncertainty, the observation of a minimum of the energy gap value at
about 60$^{\text{o}}$ qualitatively supports our model.
\begin{figure}[ht]
\centerline{
\includegraphics[width=7.2cm,clip=]{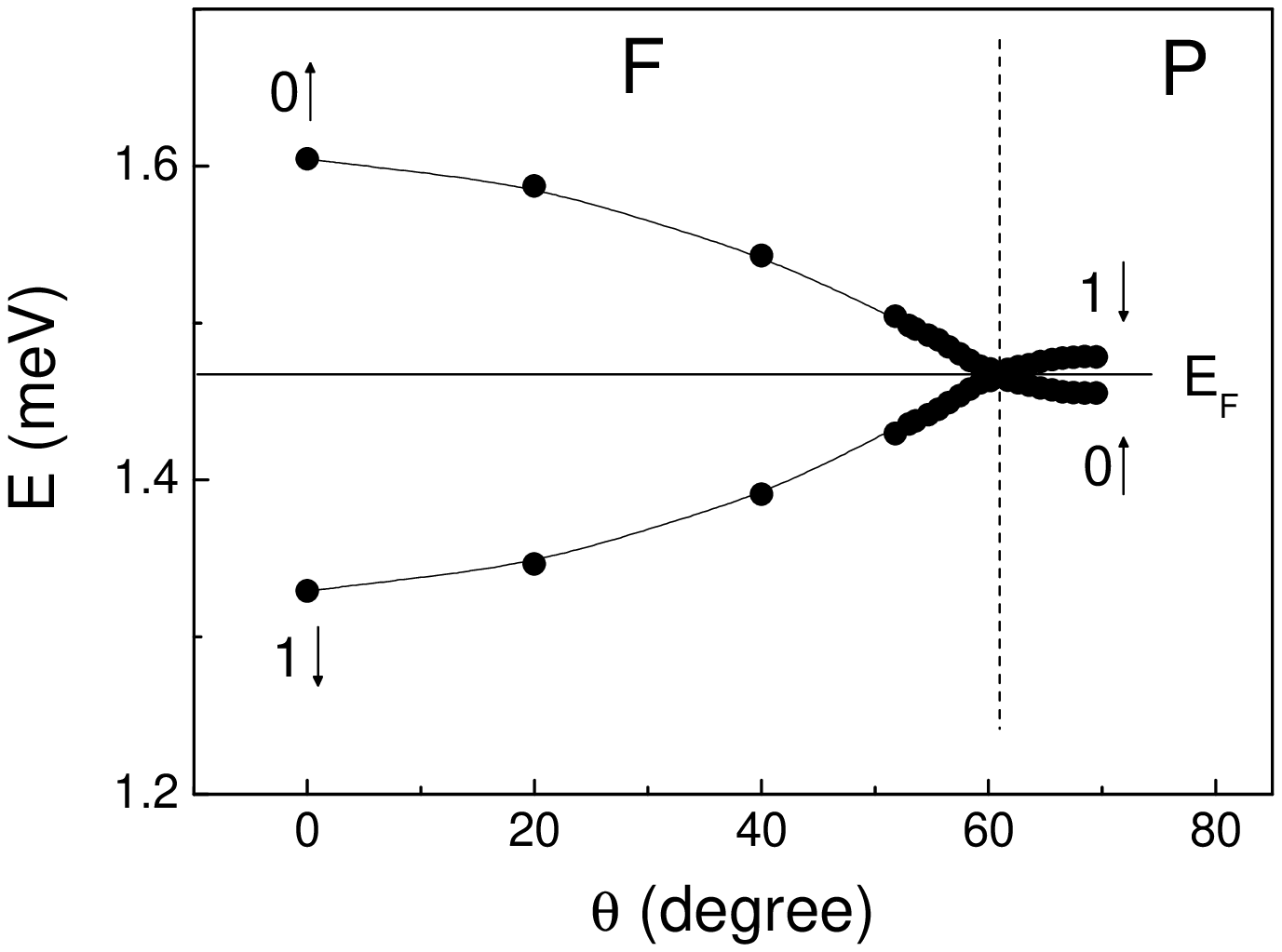}
} \caption{Energies of the LLs 0$\uparrow$ and 1$\downarrow$ vs.
angle $\Theta$ for the sample with
$p$=2$\times$10$^{11}$\,cm$^{-2}$.
 \label{fig:Levels3117}}
\end{figure}

These facts confirm that the observed F-P transition is indeed
 associated with the crossing of the LLs 0$\uparrow$
and 1$\downarrow$ at 60$^{\textrm{o}}$. Now, knowing the activation
energy dependence on $\Theta$ and using the value $\Delta E$=0.28
meV found in Ref.~\onlinecite{7} for $\Theta$=0, we can get a more
accurate angle dependence of the energies of the levels 0$\uparrow$
and 1$\downarrow$. It is presented in Figure~\ref{fig:Levels3117}.

The F-P transition is expected to be accompanied by the formation of
ferromagnetic domains.  According to Ref.~\onlinecite{9}, the domain
formation should be manifested in an anisotropy of the
magnetoresistance, i.e. in a tilted field the value of the
magnetoresistance should depend on the orientation of
$B_{\parallel}$, the in-plane projection of the magnetic field, with
respect to the current.  For example, an anisotropy in the region
where LLs cross has been reported in several papers for
GaAs/AlGaAs~\cite{10} and n-Si/SiGe~\cite{11,12} heterostructures.

\begin{figure}[ht]
\centerline{
\includegraphics[width=9cm,clip=]{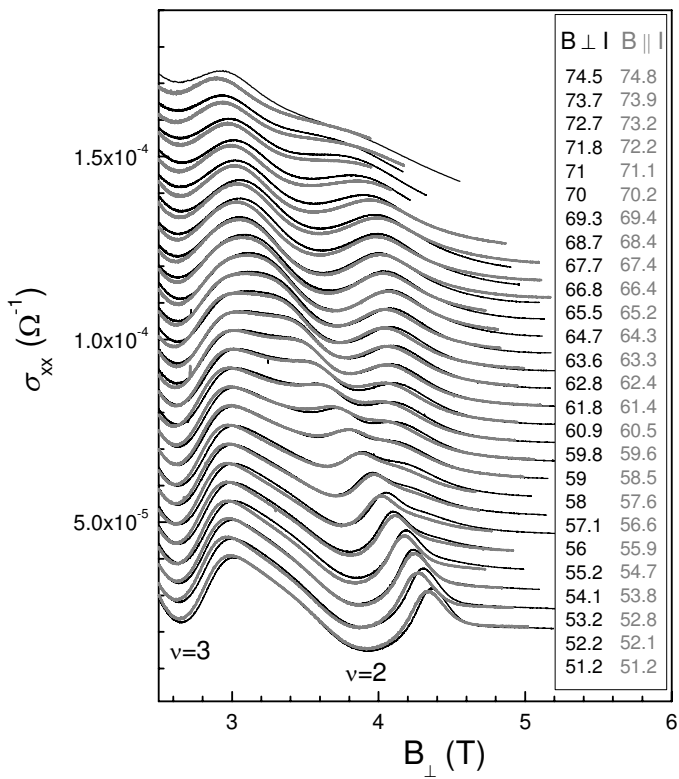}
} \caption{Dependences of the $\sigma_{xx}$ on the normal component
of the magnetic field for different tilt angles shown for two
orientations of the magnetic field $B_{\parallel} \parallel I$ and
$B_{\parallel} \perp I$ at $T=0.3$ K. The curves for each angle are
shifted by 5$\times$10$^{-6}$ $\Omega^{-1}$ for clarity.
 \label{fig:sxxAnis}}
\end{figure}

We tilted the sample in the two possible orientations, keeping the
field projection ($B_{\parallel} \parallel I$) parallel and
($B_{\parallel} \perp I$) perpendicular to the current, but did not
observe any anisotropy of the magneto-resistance in the vicinity of
the transition.  Figure~\ref{fig:sxxAnis} illustrates the dependence
of the conductivity on the normal component of the applied magnetic
field $B_{\perp}$ at different angles and for both orientations of
the in-plane projection of $B$ relative to the current.

As seen in Figure~\ref{fig:sxxAnis}, the curves for the different
directions of the in-plane projection of the magnetic field
($B_{\parallel}
\parallel I$
and $B_{\parallel} \perp I$) virtually coincide, i. e. in our case
the anisotropy of the conductivity is absent with a high degree of
accuracy.

\begin{figure}[ht]
\centerline{
\includegraphics[width=7.2cm,clip=]{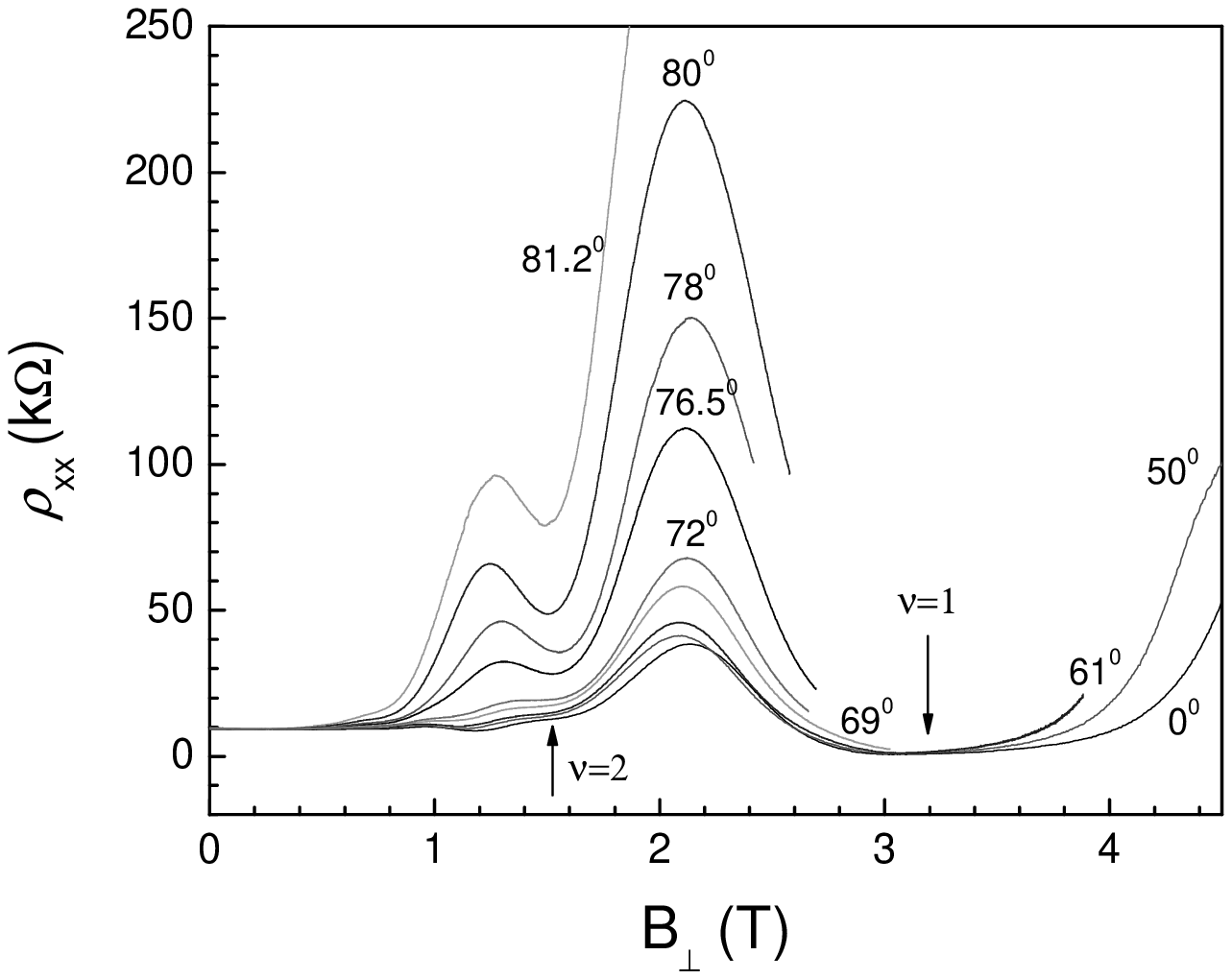}
} \caption{Dependences of the $\rho_{xx}$ on the normal component of
the magnetic field for different tilt angles. $T$=0.2 K.
 \label{fig:rxx3212}}
\end{figure}

We also carried out similar studies at $T$=(18 - 200) mK for the
lower density p-Si/SiGe/Si sample with $p=7.2\times 10^{10}$
cm$^{-2}$.  The dependence of the resistivity  $\rho_{xx}$ on the
magnetic field for different tilt angles are shown in
Figure~\ref{fig:rxx3212}.  We particularly notice that, at tilt
angles $\Theta>$ from $0^{\text{o}}$ to 70$^{\text{o}}$, the
oscillations corresponding to $\nu$=2 are extremely weak.  They only
start manifesting themselves for $\Theta>$ 70$^{\text{o}}$. At
$\nu$=3/2, a maximum of resistance appears similar to the one we
observed in the other sample, with a magnitude that depends strongly
on the tilt angle.

\begin{figure}[ht]
\centerline{
\includegraphics[width=7.2cm,clip=]{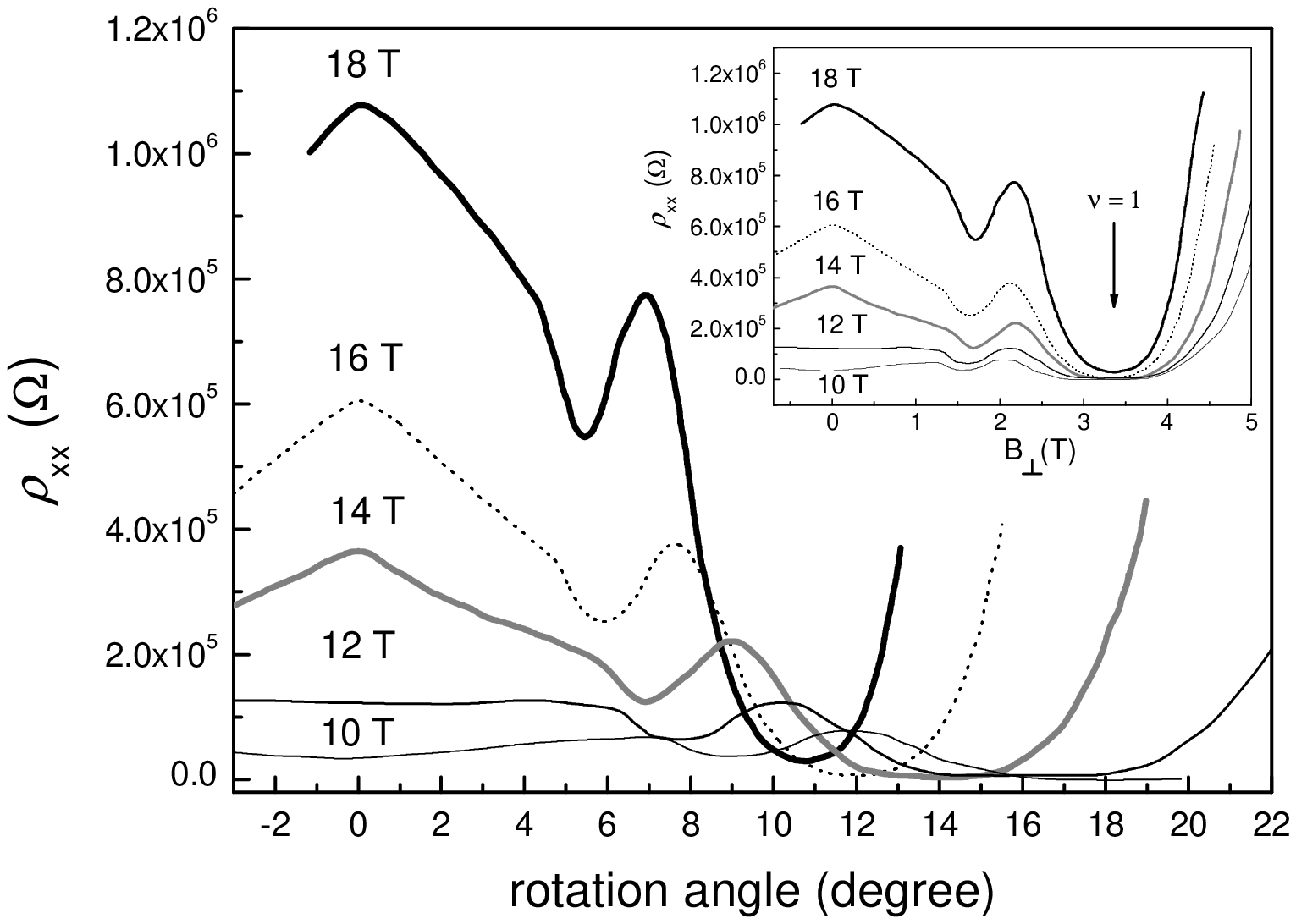}
}\caption{Resistance  $\rho_{xx}$ as a function of the field tilt
angle with respect to the plane of the 2D layer at different values
of the total magnetic field, $T \approx$0.4 K.  Inset: $\rho_{xx}$
as a function of the normal component of the magnetic field $B$.
 \label{fig:rxxRotate}}
\end{figure}

Yet the oscillations at $\nu$=2 are clearly visible in another way
of measuring the magnetoresistance: when the sample is rotated in a
fixed total magnetic field, the perpendicular field component
$B_{\perp}$ causes oscillations at the angles determined by the
concentration of charge carriers in the sample.
Figure~\ref{fig:rxxRotate} shows such an angle dependence of the
magnetoresistance measured at several fixed magnetic fields, where
the oscillation related to $\nu$=2 can be seen to move from a tilt
of about 9$^{\text{o}}$ at 10 T to 5$^{\text{o}}$ at 18 T.  This
corresponds to $B_{\perp}$= 1.7 T in each case, as shown in the
Figure~\ref{fig:rxxRotate} inset.~\cite{11} The field value for
$\nu$=2 $B$=1.7 T is slightly different from data shown above. This
is probably a result of an ageing of the sample as the experiments
of Ref.~\onlinecite{11} were done much earlier.

\begin{figure}[ht]
\centerline{
\includegraphics[width=9cm,clip=]{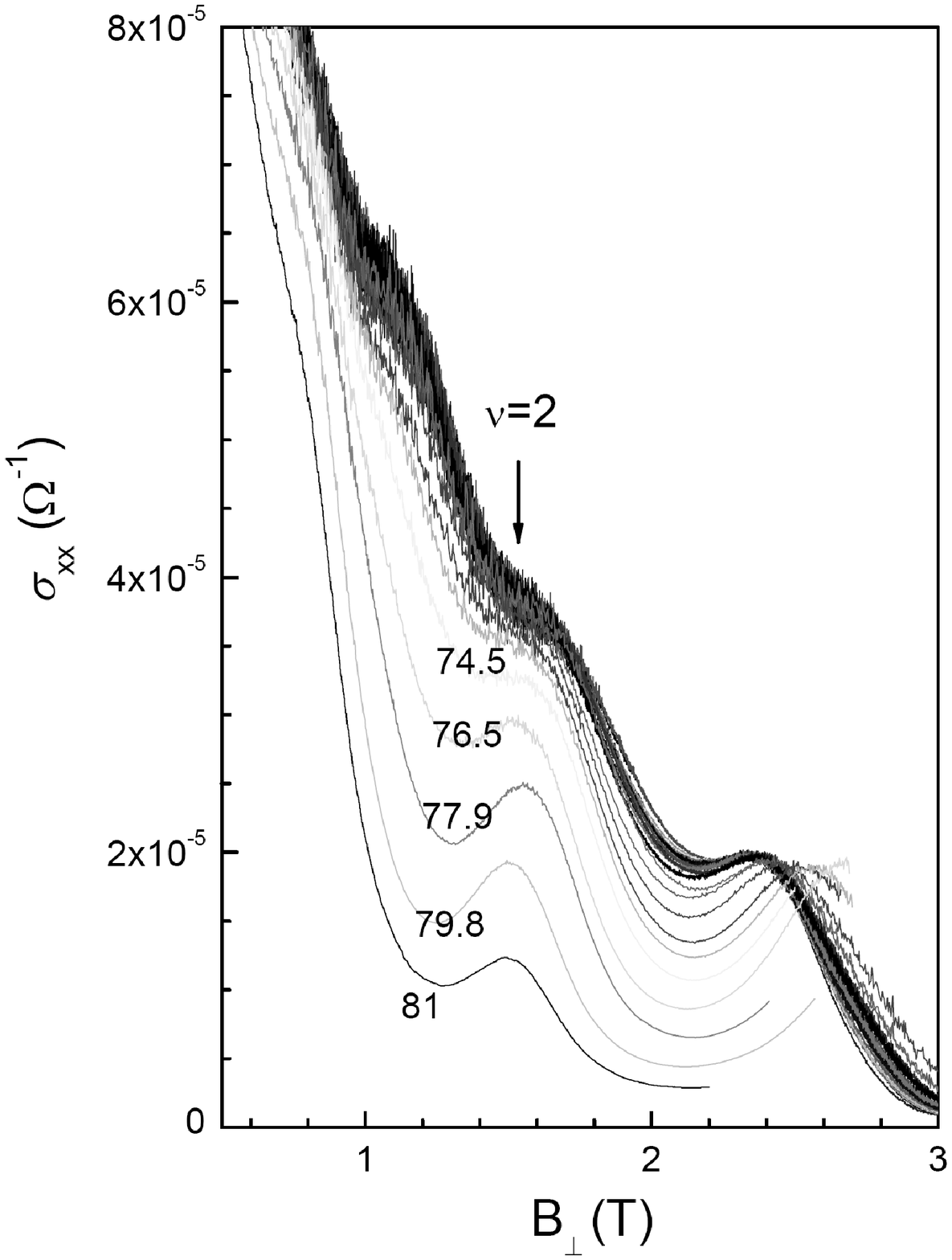}
} \caption{Dependences of the $\sigma_{xx}$ on the normal component
of the magnetic field for different tilt angles for the sample
$p=7.2 \times 10^{10}$ cm$^{-2}$; $T=0.2$ K.
 \label{fig:sxx3212}}
\end{figure}

The dependence of the conductivity $\sigma_{xx}$ on the normal
component of the magnetic field $B_{\perp}$ is shown in
Figure~\ref{fig:sxx3212} at different tilt angles, with
$B_{\parallel} \parallel I$. Since the oscillations of $\rho_{xx}$
at high tilt angles are observed against a background of high
resistance with  $\rho_{xx} \gg  \rho_{xy}$, it turns out that
$\sigma_{xx} \sim 1/\rho_{xx}$, so minima in $\rho_{xx}$ correspond
to maxima in $\sigma_{xx}$, as observed at $B_{\perp}\approx$1.5 T
in Figure~\ref{fig:sxx3212}.

The absence of oscillations at magnetic fields corresponding to
$\nu$=2 in the range of angles (0-70)$^{\text{o}}$ indicates that
the 0$\uparrow$ and 1$\downarrow$ LLs coincide.  The appearance of
these oscillations for $\Theta > $70$^{\text{o}}$ is due, in our
opinion, to the fact that the levels begin to diverge, resulting in
the energy gap opening up.  Apparently, the gap opening in the
sample with $p=7.2 \times 10^{10}$cm$^{-2}$ is associated with the
angle dependence of the g-factor.  The g-factor in this material is
anisotropic~\cite{1} and depends on the magnetic field tilt angle
relative to the sample surface normal.  If the g-factor had an axial
symmetry we could write $g^*=\sqrt{g_{\perp}^2
\cos^2(\Theta)+g_{\parallel}^2 \sin^2(\Theta)}$, where $g_{\perp}$
is the g-factor with the magnetic field perpendicular to the 2DHG,
and $g_{\parallel}$ is with the magnetic field parallel to the 2DHG.
For strong anisotropy, when $g_{\parallel}$=0 (as it should be in
our structure) this reduces to $g^* = g_{\perp} \cos \Theta$.
However, if such a dependence of the g-factor were to occur, then
the F-P transition should not be observed.

Unfortunately, we are unable to make reliable calculations and
determine the width of the gap appearing in the sample with $p=7.2
\times 10^{10}$cm$^{-2}$ due to the large magnetoresistance produced
by the parallel magnetic field in this sample.~\cite{13}  It should
be noted that the values of  $\rho_{xx}(B)$ and  $\sigma_{xx}(B)$,
on which background the oscillations develop, strongly depend on the
magnetic field, and the greater the angle   the stronger is this
dependence. So, it does not seem to be possible to reliably separate
the small oscillations at $\Theta > $70$^{\text{o}}$ from the smooth
background of $\rho_{xx}(B)$, which is about 10$^6$ ohms.  (Such
problem for the sample with $p=2 \times 10^{11}$cm$^{-2}$ did not
arise because the overall change $\rho_{xx}(B)$/$\rho_{xx}$(0) in a
parallel magnetic field of 18 T did not exceed a factor of 4, and
the in-plane resistance was only about 10$^4$ ohms).

Thus, the complete F-P transition in the sample with $p=7.2 \times
10^{10}$cm$^{-2}$ is not observed in tilted fields.  In a wide range
of angles $\Theta$=(0-70)$^{\text{o}}$ the 0$\uparrow$ and
1$\downarrow$  LLs are still coinciding, and only for $\Theta
>$70$^{\text{o}}$ is
there a gap in the hole energy spectrum arising as a result of a
divergence of the LLs.

\section{Conclusion}

The ferromagnetic-paramagnetic transition is observed in a
p-Si/GeSi/Si sample with $p=2 \times 10^{11}$cm$^{-2}$ at a magnetic
field corresponding to filling factor $\nu\approx$2.  It appears as
a result of a change in the relative position of the 0$\uparrow$ and
1$\downarrow$  LLs as a function of the tilt angle $\Theta$.  This
fact was first demonstrated in Ref.~\onlinecite{7} and is confirmed
in this paper by measurements of the energy gap dependence on the
angle $\Theta$. For this sample we also demonstrate an absence of
anisotropy of  xx with respect to the magnetic field projection on
to the sample plane, despite such an anisotropy having been proposed
in Ref.~\onlinecite{9}.  At the same time, in the sample with $p=7.2
\times 10^{10}$cm$^{-2}$ the ferromagnetic-paramagnetic transition
is not observed. In a wide range of angles $\Theta =
$0-70$^{\text{o}}$ the LLs 0$\uparrow$ and 1$\downarrow$  coincide,
and only for $\Theta > $70$^{\text{o}}$ does a gap open in the hole
spectrum as a result of the LLs diverging.

Ambiguity in the results observed by various
authors~\cite{1,2,3,4,5,6}, as well as ourselves, on different
p-Si/GeSi/Si samples is due, in our opinion, to dissimilar
dependences of the g-factors on the magnetic field tilt angle.  This
is caused by different levels of disorder in all these samples,
since disorder can lead to breaking of the axial symmetry.

\subsection{Acknowledgments}

The authors are grateful to E. Palm, T. Murphy, J.H. Park, and G.
Jones for their help with the experiments. This work was supported
by grant of RFBR 11-02-00223, grant of the Presidium of the Russian
Academy of Science, the Program "Spintronika" of Branch of Physical
Sciences of RAS. The NHMFL is supported by the NSF through
Cooperative Agreement No. DMR-0654118, the State of Florida, and the
 US Department of Energy.

\vfill\eject


\begin{thebibliography}{25}

\bibitem{1} S.I. Dorozhkin, Pis'ma v ZhETF \textbf{60}, 578 (1994);
[JETP Lett. \textbf{60}, 595 (1994)].

\bibitem{2} S.I. Dorozhkin, C.J. Emeleus, T.E. Whall,
G. Landwehr, and O.A. Mironov,  Pis'ma v ZhETF \textbf{62}, 511
(1995); [JETP Lett. \textbf{62}, 534 (1995)].

\bibitem{3} P.T. Coleridge, A.S. Sachrajda,
P. Zawadzki, R.L. Williams, and H. Lafontaine,
Sol. State Commun. \textbf{102}, 755 (1997).

\bibitem{4} M.R. Sakr, Maryam Rahimi,
S.V. Kravchenko, P.T. Coleridge, R.L. Williams, and J. Lapointe,
Phys. Rev. B \textbf{64}, 161308 (2001).

\bibitem{5} P.T. Coleridge, Sol. State Commun. \textbf{127}, 777 (2003).

\bibitem{6} R.B. Dunford, E.E. Mitchell, R.G. Clark,
V.A. Stadnik, F.F. Fang, R. Newbury, R.H. McKenzie, R.P. Starrett,
P.J. Wang, and B.S. Meyerson, J. Phys.: Condens. Matter \textbf{9},
1565 (1997).


\bibitem{7}  I.L. Drichko, I.Yu. Smirnov, A.V. Suslov, O.A. Mironov,
and D.R. Leadley, ZhETF \textbf{138}, 557 (2010); [JETP
\textbf{111}, 495 (2010)].


\bibitem{8} A.J. Daneshvar, C.J.B. Ford, M.Y. Simmons, A.V. Khaetskii, A.R. Hamilton, M. Pepper, and D.A. Ritchie, Phys. Rev. Lett.
\textbf{79}, 4449 (1997).

\bibitem{9}  J.T. Chalker, D.G. Polyakov, F. Evers, A.D. Mirlin, and P. Wolfle, Phys. Rev. B \textbf{66}, 161317 (2002).

\bibitem{10} W. Pan, H.L. Stormer, D.C. Tsui, L.N. Pfeiffer, K.W. Baldwin,
and K.W. West, Phys. Rev. B \textbf{64}, 121305(R) (2001).

\bibitem{11} U. Zeitler, H.W. Schumacher, A.G.M. Jansen, and R.J.
Haug, Phys. Rev. Lett. \textbf{86}, 866 (2001).

\bibitem{12} Kiyohiko Toyama, Takahisa Nishioka, Kentarou Sawano, Yasuhiro
Shiraki, and Tohru Okamoto, Phys. Rev. Lett. \textbf{101}, 016805
(2008).

\bibitem{13} I. L. Drichko, I. Yu. Smirnov, A. V. Suslov, O. A. Mironov, and D. R. Leadley, Phys. Rev. B \textbf{79}, 205310 (2009)

\end{thebibliography}
\end{document}